# Constructing longulence in the Galerkin-regularized nonlinear Schrödinger and complex Ginzburg-Landau systems

Jian-Zhou Zhu (朱建州)

[a]*Su-Cheng Centre for Fundamental and Interdisciplinary Sciences Gaochun Nanjing 211316 China*

**Abstract**

(Quasi-)periodic solutions are constructed analytically for Galerkin-regularized or truncated nonlinear Schrödinger (GrNLS) systems preserving finite Fourier freedoms. GrNLS admits travelling-wave or multi-phase solutions, including monochromatic solutions independent of the truncation and quasi-periodic ones with or without additional on-torus invariants. Numerical tests show that instability leads such solutions to nontrivial longulent states with remarkable solitonic structures (called "longons") admist disordered weaker components, corresponding to presumably whiskered tori. In the strong-coupling limit (e.g., the self-phase modulation equation in optics), neutral stability holds for the condensates, without the modulational instability, but not generally for other multi-phase (quasi-)periodic solutions from some of which the longulent state developed is also adressed. The possibility of nontrivial Galerkin-regularized complex Ginzburg-Landau longulent states is also discussed for motivation.

## 1. Introduction

A wide spectrum of multidisciplinary processes in Nature, ranging from hydrodynamics, plasma physics, optics to Bose-Einstein condensates (BEC), can be well modeled by the nonlinear Schrödinger (NLS) or Gross-Pitaevskii (GP) equation. In the transition from classical soliton theory to its quantum counterpart, particularly in the Hamiltonian framework, the NLS equation is often regarded as simpler and more fundamental than the Korteweg-de Vries (KdV) equation [1]. The two models also present different nonlinear physics, with no KdV but possible NLS finite-time or asymptotic blow-up (c.f., Ref. [2] and references therein for the focusing cubic case focused here.)

We briefly introduce the most basic, in terms of nonlinear mathematical physics, and closely relevant (to solitonic structures) backgrounds in the following. Concerning our periodic problem, associated to NLS, probably the most remarkable phenomena are the fractalization and quantum revival, associated to the Talbot effect ([3, 4, 5]), and solitons (see, e.g., Refs. [6, 7] and references therein for traditional Hamiltonian theory and most recent developments on soliton gas.) For the spectral theory of similar systems integrable by inverse scattering transform, see, e.g, Ref. [8] and references therein for inifinte-gap theory, and, Ref. [9] and references therein for recent advancements of unified transform/Fokas method over the classical inverse scattering transform for periodic problems. The classical studies of periodic and quasi-periodic NLS solutions can be found in, e.g., Refs. [10, 11, 12]. Finally, related to the instability in our numerical tests, note that, even for the infinite-line problem, the modulational-instability-stage problem is nontivial (see, e.g., the recent different results of Zakharov-Gelash [13] and Biondini-Mantzavinos [14].)

Other ingenious methods have been used to study soltion systems, but in general the so-called nonintegrable equations are in lack of an effective systematic theory of solutions; for instance, the understanding of nonintegrable soliton gases (e.g., NLS with nonlinearity of orders higher than 3 [15]) heavily relies on the help of numerical investigations. Of course, "integrability", byond that of the Liouville sense and that associated to the inverse scattering method, by itself is neither completely defined in mathematics nor physically

*Email address:* jz@sccfis.org (Jian-Zhou Zhu (朱建州) )

well understood. We need a working framework for general conservative systems towards which this work belongs to the efforts (see also the simultaneous communications [16, 17]). The most closely relevant models, at least formally, are of course the corresponding Galerkin-regularized (Gr) systems preserving finite Fourier freedoms (with wavenumber modules, $|k| \leq K$) and important parts of the mode interaction structure, including some conservation laws [18, 16]. [For the real variable $u$ dealt with elsewhere [16], conjugate Fourier coefficients $\hat{u}_{\pm k}$ work together to form a "mode"; now, the (Gr)NLS variable is complex, with independent "modes" of wavenumbers $\pm k$. So, to avoid confusion, we will try to resist using such a notion.] Although the Lax pair structure is absent, we may recover and extend certain elements, say, by combining Fourier expansion with truncation and exploring other analytical approaches.

Paralleling the studies of (quasi-)periodic patterns in hydrodynamic-type Gr-systems, such as the Burgers-Hopf (BH), compacton-and-peakon (CP), and KdV equations [16], and, the untruncated even-odd alternating Korteweg-de Vries (aKdV) equation [17], we will explore what is new in the mathematical physics of the Gr-system with higher-order nonlinearity for the complex order parameter (here cubically nonlinear GrNLS or GrGP).

The common features of various Gr-systems, such as the GrNLS to be elaborated, include

- the loss of the some of the invariants, sometimes infinitely many (and the integrability) for particular systems,
- leaving a few "rugged" ones, the regularization of the structures (singular for BH and CP),
- and the admittance of new travelling waves, (quasi-)periodic orbits and, as will be explained, "pseudo-periodic" (statistically) stable "longulent state" or "longulence".

Longulence is characterized by solitonic structures, that we call "longons", accompanied by the less-ordered components, corresponding to a presumably whiskered torus whose stable manifold is responsible for the solitonic structures with apparent periodic character (probably quasi- or almost-periodic) and whose unstable manifold for the chaotic components: the orbit can somehow escape with perturbations from the unstable "whiskers" and come back to the stable manifold. We have to introduce a new term, "pseudo-periodicity", for the combination of periodicity (of the solitonic longons) and the chaoticity (of the disordered components).

More explanation of the choice of the term "longon" follows. First of all, solitonic structures might all be loosely called "solitons" which however got some mathematically rigorous meaning (in the spectral theory asociated to the inverse scattering method) for which even the original Zabusky-Kruskal "solitons" [19] are not strictly precise. More importantly, the objective solitonic structures studied here are quite universal in a variety of Gr-systems (including a series of hydrodynamic-type ones in Ref. [16]) that we have investigated, deserving a particular name, for distinguishment and for convenience; and, they appear also unique, in the sense of, for instance, being always accompanied by disordered weaker components, which is somehow echoeing with the spirit of "Long" [i.e., the oriental dragon or "龙" in Chinese Pinyin, embodying power and upholding order amidst (potential) chaos; "Long" is also the Chinese zodiac for the year 2024 when the main parts of the work were completed.]

Since the GrNLS system admit typical monochromatic solutions/condensates independent of the truncation (and stable in the strong-coupling limit or the self-phase modulation case without the linear-dispersion term) and since in suitable situations GrNLS can well approximate the full/untruncated NLS in the sense of convergence to the latter (as is the case in Ref. [20] but not for general Gr-systems such as the above-mentioned GrBH discussed in Ref. [16] and references therein), the following clarification should be made to avoid possible confusion in later discussions: here we require the

*nontrivial* longulent state or longulence be also with clear truncation effects; so, neither the state without a disordered component nor that with solitonic structure(s) but also already with convergence to the full-/untruncated-system one, with, say, soliton turbulence (see, e.g., recently Ref. [7] and references therein), satisfies this criterion. We can of course say NLS soliton turbulence is a special kind of longulence, or the limit of longulence with $K \to \infty$, which case however is not of our interest here for the truncation effects.

Unlike those hydrodynamic-type systems in Ref. [16] where torus-specific or on-torus invariants (varying outside the torus, thus not rugged) have to be introduced to construct quasi-periodic solutions to support the *a-posteriori* Kolmogorov-Arnold-Moser (KAM) argument for the longulent states, we will see that GrNLS rugged invariants are already sufficient to augment quasi-periodic orbits. There are other important differences. For instance, no-linear-dispersion NLS, or more precisely (for actually no direct connection to Schrödinger) the strong-coupling limit, i.e., the "self-phase modulation" equation in optics, also admits monochromatic-wave solutions (without modulational instability!)

The complex Ginzburg-Landau (CGL) equation is coefficient-complexified, or, with small imaginary part(s) in the coefficient(s), a perturbation resulting in damping and (autonomous) driving to NLS (thus, among others, the interesting *persistence* problem [21]). Actually, in our context, CGL also supports solitonic structures of great physical importance (e.g., recently the application of Kerr and Nozaki-Bekki solitons in optics [22]) and presumably high-dimensional whiskered tori [23, 24], with also chaotic dynamics somewhat trackable [25] and mimicing aspects of fluid turbulence [26]. It is then more helpful to consider GrNLS in the broader context of GrCGL [27, 28, 29, 30]. The problem of GrCGL appears much more challenging due to the fact that the forcing is nonlinearly dependent on the order parameter itself, but preliminary analyses and numerical tests will be offered for comparison and motivation. So far, not as expected, we have not found evidence of the persistence of the GrNLS longulence with respect to the GrCGL perturbation, which however does not necessarily exclude the possibility. So, due to the difficulty, this work is not formally as complete as that for hydrodynamic-type Gr-systems [16], although we can introduce other independent driving and damping into NLS, like those with respect to which the GrBH longulence is explicitly shown to be persistence: with the very different properties of the objective models and distinct results, these two independent articles, somehow complementing each other, are respectively written in a self-contained way, both involving notions such as longons and longulence though.

Using GrNLS or GrGP with sufficient number of the eigen modes of the harmonic oscillator potential to probe the quasi-integrability of the full dynamics [20] and the higher-dimensional GrGP thermalization aspect (e.g., Ref. [31] and references therein) belong to different lines of research. Our distinct work is organized as follows. Sec. 2 finds analytically the exact GrNLS travelling-wave and (quasi-)periodic multi-frequency solutions, emphasing the critical sets specified by rugged and torus-specific invariants; Sec. 3 discovers the universal longulent states numerically, with remarks including preliminary considerations on the (non)persistence of GrNLS longons against the GrCGL perturbation; and, finally, Sec. 4 naturally extends the discusion of GrCGL itself, including the expectation of quasi-periodic tori and longulence, and, the challenge to construct them.

## 2. The problem and solutions

We start with the Hamiltonian formulation [1] of the $2\pi$ periodic NLS problem directly in the Fourier ($k$) space, as Gardner did for KdV [32], deferring the Poisson structure in physical ($x$) space to the point when needed.

Let $\Psi(x,t) = \sum_n \hat{\Psi}_n(t) e^{\hat{i}nx}$ (where $\hat{i}^2 = -1$) with $2\pi$ $x$-period solve the NLS equation (2.3) below with $g = 0$, we have

$$\dot{q}_k = \frac{\partial \mathcal{H}}{\partial p_k},\ \dot{p}_k = -\frac{\partial \mathcal{H}}{\partial q_k}, \tag{2.1}$$

with $q_k = \hat{\Psi}_k$, $p_k = \hat{i}\hat{\Psi}^*_k$ for the Fourier coefficient $\Psi_k$ and its conjugate $\hat{\Psi}^*_k = (\widehat{\Psi^*})_{-k}$ of each wavenumber $k$, and,

$$\mathcal{H} = \sum_n \left( n^2 |\hat{\Psi}_n|^2 \mp \sum_{k+l-j=n} \hat{\Psi}_k \hat{\Psi}_l \hat{\Psi}^*_j \hat{\Psi}^*_n \right). \tag{2.2}$$

The upper sign ("$-$" here) is for the focusing case, and the lower ("$+$" here) defocusing: we will eventually focus on the focusing case.

For Galerkin-regularized or Fourier-truncated $\psi = \sum_{|n|\leq K} \hat{\psi}_n(t) e^{\hat{i}nx}$ ($\leftrightarrows \Psi$, through a simple "hard cutoff" as a pseudo-differential operation in the language of analysis) and "well-prepared" initial data $\psi(0)$ with $\hat{\psi}_m(0) = 0\ \forall m > K$ [1], the GrNLS system involves a Galerkin function/force $g$ with the effect of projecting the dynamics on to the space of $|k| \leq K$,

$$\hat{i}\partial_t\psi + \partial_{xx}\psi \pm 2|\psi|^2\psi = g; \tag{2.3}$$

$$\hat{i}\dot{\hat{\psi}}_n - n^2\hat{\psi}_n \pm 2 \sum_{k+l-j=n} \hat{\psi}_k\hat{\psi}_l\hat{\psi}_j^* = \hat{g}_n : \tag{2.4}$$

$$\hat{g}_m = \begin{cases} \pm 2 \displaystyle\sum_{k+l-j=m} \hat{\psi}_k\hat{\psi}_l\hat{\psi}_j^* & \text{for } K < |m|, \\ 0 & \text{otherwise.} \end{cases} \tag{2.5}$$

The explicit expression of $g$ in the GrBH case was already used by Tadmor [33] for mathematical estimation.

It is seen that the Hamiltonian formulation through Eq. (2.2), with $\mathcal{H} \leftrightarrows$

$$H = \sum_{|k|\leq K} \left( k^2|\hat{\psi}_k|^2 \mp \sum_{i+l-j=k}^{|i|,|l|,|j|\leq K} \hat{\psi}_i\hat{\psi}_l\hat{\psi}_j^*\hat{\psi}_k^* \right),$$

still applies, similar to the KdV case [32], and the invariants,

$$M_{-1} = \sum_{|k|\leq K} |\hat{\psi}_k|^2 \text{ and } M_0 = \sum_{|k|\leq K} k|\hat{\psi}_k|^2, \tag{2.6}$$

are still conserved, defining two other Hamiltonian flows with $\hat{i}\partial_t\psi = \delta M_\tau/\delta\psi^*$, respectively for $\tau = -1$ and 0: Gardner [32] actually used the finite-mode case for the intermediate stage in proving some results — see Ref. [18] for physical-space analysis, or, Ref. [34] for a direct calculation of Fourier mode interactions to show the preservation of GrBH $H$. No other NLS invariants are known to be preserved by GrNLS. Such (non)conservation laws can be argued similarly to the analysis for GrBH in Ref. [18], but it is actually more straightfrowardly seen in $k$-space, following Gardner [32], which is one of the reason for us to emphasize the $k$-space formulation in the above.

Note that, unlike the quadratic interaction with $\hat{g}_m \equiv 0$ for $|m| > 2K$ [16], we now have $\hat{g}_m \equiv 0$ for $|m| > 3K$.

### *2.1. One- and two-frequency/phase solutions*

With the cubic nonlinearity, a remarkable apparent difference between the current quartic interaction to the triadic one for the quadratically nonlinear systems is that a $\hat{\psi}_k$ can interact with itself without involving $\hat{\psi}'_k$ of $k' \neq k$. So, the simplest nontrivial/nonzero GrNLS solution to Eq. (2.4) is that occupying only a single wavenumber $k = S$ and satisfying

$$\hat{i}\dot{\hat{\psi}}_S = S^2\hat{\psi}_S \mp 2|\hat{\psi}_S|^2\hat{\psi}_S, \tag{2.7}$$

also solving the untruncated NLS, the well-known monochromatic wave or condensate. Assuming $\hat{\psi}_S = Ae^{\hat{i}\theta_S(t)}$ with constant $A = \alpha S$, we get

$$\hat{\psi}_S = Ae^{-\hat{i}(1\mp 2\alpha^2)S^2 t} \text{ or } \psi = Ae^{\hat{i}S[x-(1\mp 2\alpha^2)St]} \tag{2.8}$$

where $\theta_S(0) = 0$ is taken for simplicity: such focusing GrNLS waves become stationary for $\alpha = 1/\sqrt{2}$.

[1] "Ill-prepared" initial data $\psi(0)$, $\exists m > K$, $\hat{\psi}_m(0) \neq 0$ are not of our interest here. Also, we resist introducing further regularizations, such as the truncations on the quadratic interactions $|\psi\psi|^*$ or $\psi^2$ before forming $|\psi|^2\psi^*$.

In the case of no linear-dispersion regularization (NLDNLS), i.e., the strong-coupling limit or the self-phase modulation equation (in optics)

$$\hat{i}\partial_t \Psi \pm 2|\Psi|^2\Psi = 0$$

and the corresponding Gr-version, the solution (2.8) changes accordingly to $\hat{\psi}_S = Ae^{\pm 2\hat{i}\alpha^2 S^2 t}$ or $\psi = Ae^{\hat{i}S[x \pm 2\alpha^2 St]}$. [Without the linear dispersion, there is no direct connection with Schrödinger, but in our context it is convenient to still denote the model and the Gr-version with NLDNLS and GrNLDNLS.] The situation looks similar to the (Gr)BH case, but, unlike the latter, monochromatic wave solution solves both NLDNLS and GrNLDNLS. Also, straightforward linear analysis shows that such (Gr)NLDNLS condensates are neutrally stable, without the modulational instability in the corresponding (Gr)NLS case. Such stability however is not generic for other more general GrNLDNLS multi-phase (quasi-)periodic solutions whose instability leads to longulent states.

Extremizing GrNLS Hamiltonian $H$ constrained by $M_{-1}$ and $M_0$, with the respective (real) lagrangian multiplier $-\lambda_{-1}$ and $-\lambda_0$, leads to

$$\frac{\delta H}{\delta \psi^*} - \lambda_{-1}\frac{\delta M_{-1}}{\delta \psi^*} - \lambda_0 \frac{\delta M_0}{\delta \psi^*} = 0 : \tag{2.9}$$

$$\hat{i}\dot{\hat{\psi}}_k = k^2\hat{\psi}_k \mp 2 \sum_{j+l=m+k}^{|k|,|m|,|l|,|j|\le K} \hat{\psi}_j\hat{\psi}_l\hat{\psi}^*_m = (\lambda_{-1} + \lambda_0 k)\hat{\psi}_k, \tag{2.10}$$

resulting in the solutions $\hat{\psi}_k = \hat{\psi}_{k0}e^{-\hat{i}(\lambda_{-1}+\lambda_0 k)t}$. Initial $\hat{\psi}_{k0}$ is determined by the equality of the middle and right-hand sides, while the time dependence by the left- and right-hand sides. Such a special-solution approach resembles the integrable cases for finite-band solutions [12, 35, 36].

Travelling waves may be realized by Eq. (2.10) with $\lambda_{-1} = 0$. When occupying multiple wavenumbers, the solutions with $\lambda_{-1} \neq 0$ to Eq. (2.10), or even with more general $\omega_K(k)\hat{\psi}_k$, with $\omega_K$ being, say, higher-order polynomials of $k$, on the right hand side, are generally not travelling waves. We now consider solutions occupying only wavenumbers of $k = \pm S$.

Our analyses for focusing and defocusing cases are formally the same. So, from now on we restrict ourselves to the focusing case, leaving the possible other interesting aspect of defocusing GrNLS aside for the time being.

Then, because the data occupying $k = \pm S$ excite other wavenumbers only of $k = \pm 3S$, we obtain, for instance, the solutions occupying $k = \pm S$ for $S \le K \le 3S-1$, with $|\hat{\psi}_{\pm S}|^2 = (S^2 \pm 3\lambda_0 S - \lambda_{-1})/6$. The latter is not always guaranteed to be nonnegative by arbitrary combinations of $\pm S$, $\lambda_{-1}$ and $\lambda_0$, thus indicating nontriviality of the existence of such solutions specifically, and of those occupying more wavenumbers in general. The solutions read:

$$\psi = \frac{\sqrt{S^2 + 3\lambda_0 S - \lambda_{-1}}}{\sqrt{6}} e^{\hat{i}[Sx - (\lambda_{-1}+\lambda_0 S)t + \theta_+]} + \frac{\sqrt{S^2 - 3\lambda_0 S - \lambda_{-1}}}{\sqrt{6}} e^{-\hat{i}[Sx + (\lambda_{-1}-\lambda_0 S)t + \theta_-]}. \tag{2.11}$$

Such $\psi$, composed of two travelling-wave components with initial phases $\theta_\pm$, by itself is not for travelling waves in general, being quasi-priodic when $\lambda_{-1}$ and $\lambda_0$ are rationally independent/incommensurate. The reduction with $\lambda_0 = 0$ and $\theta_+ = -\theta_- = \theta_0$ for standing or rotating waves reads

$$\psi = \frac{2\sqrt{S^2 - \lambda_{-1}}\cos(Sx)}{\sqrt{6}} e^{-\hat{i}\lambda_{-1}t} e^{\hat{i}\theta_0} \tag{2.12}$$

which are still periodic in time.

*2.2. Additional torus-specific invariants*

In generel, we can have GrNLS exact solutions with modes occupying any amount of wavenumbers with the above mentioned $\omega_K$ containing accordingly the parameters to quantify the corresponding frequency components. We however do not know any other generic global rugged invariants for defining the critical set of some combined functional as we did in Sec. 2.1 to realized such solutions.

We can introduce torus-specific invariants, $M_\tau$, to construct such high-tori. A choice is the extension of Eq. (2.6),

$$M_\tau = \sum_{|k|\le K} k^{\tau+1}|\hat{\psi}_k|^2, \tag{2.13}$$

now with $\tau = 1$ and the associated

$$\frac{\delta H}{\delta\psi^*} - \lambda_{-1}\frac{\delta M_{-1}}{\delta\psi^*} - \lambda_0\frac{\delta M_0}{\delta\psi^*} - \lambda_1\frac{\delta M_1}{\delta\psi^*} = 0, \tag{2.14}$$

and, the right-hand side of Eq. (2.10) replaced with $(\lambda_{-1} + \lambda_0 k + \lambda_1 k^2)\hat{\psi}_k$ for a 3-phase solution set. Probably the simplest solutions are those occupying only wavenumbers $k = 0$ and $\pm S$, in which case, with further simplification by restricting to real initial $\psi$, we can solve the algebraic equation and obtain, for instance, $\hat{\psi}_k = \hat{\varphi}_k e^{-\hat{i}(\lambda_{-1}+\lambda_0 k+\lambda_1 k^2)t}$ with $\hat{\varphi}_k = 0$ except for

$$\begin{cases} \hat{\varphi}_S = \dfrac{\sqrt{2\lambda_{-1} - \lambda_0 S - \lambda_1 S^2 - S^2}}{\sqrt{30}} e^{\hat{i}\theta_0}, \\ \hat{\varphi}_0 = \dfrac{\sqrt{\lambda_{-1} + 2\lambda_0 S + 2S^2 + 2\lambda_1 S^2}}{\sqrt{10}}, \\ \hat{\varphi}_{-S} = \dfrac{\sqrt{2\lambda_{-1} - \lambda_0 S - \lambda_1 S^2 - S^2}}{\sqrt{30}} e^{-\hat{i}\theta_0}. \end{cases} \tag{2.15}$$

The phase parameter $\theta_0$ can be arbitrary. The two travelling-wave components of wavenumbers $\pm S$ interact to excite modes of $k = \pm 2S$, thus the above solution is *valid for* $|k|$ *truncated up to* $K = 2S - 1$.

Since both $M_{-1}$ and $M_0$ *Poisson commute* with $H$, resulting in vanishing Poisson bracket (see below), the above $M_1$ also Poisson commutes, through Eq. (2.14), with $\lambda_{-1}M_{-1} + \lambda_0 M_0$ and $H$ (thus invariant) on the torus. We seem to find the above solutions closer to the final longulent states, from the observation of the numerical test below. In the current context, the above torus-specific invariant bears some similarity to the "test functional" $\mathcal{F}$ of Ref. [21] for the Melnikov method, so we restate the result and remark on the differences below.

For comparison, we follow closely Ref. [21] for the symbolic convention and terminologies; see, e.g., Ref. [1] for more background on the complete theory of the infinitely many NLS invariants (or "local functionals/integrals of motion").

The evolution of a functional $\mathcal{F}$ under the NLS ($\Psi$) flow obeys

$$d\mathcal{F}/dt = \{\mathcal{F}, \mathcal{H}\} = -\hat{i}\int_0^{2\pi} \frac{\delta\mathcal{F}}{\delta\Psi}\frac{\delta\mathcal{H}}{\delta\Psi^*} - \frac{\delta\mathcal{F}}{\delta\Psi^*}\frac{\delta\mathcal{H}}{\delta\Psi}dx \tag{2.16}$$

the righ-hand side of which indicates the Poisson structure which is preserved by the GrNLS flow with the corresponding $F$ and $H$ defined by the truncated $\psi$, as mentioned earlier. $\mathcal{M}_{-1}$ and $\mathcal{M}_0$ are preserved by GrNLS for the reason similar to the KdV or Burgers-Hopf case [32, 18, 16, 34], as mentioned before. Presumably any (higher-order) NLS invariant $\mathcal{M}$ other than $\mathcal{M}_{-1}$ and $\mathcal{M}_0$ (with the corresponding $M$ redefined by $\psi$) are not supposed to be still preserved by GrNLS, also similar to KdV or Burgers-Hopf.

Now, for the tori defined by Eq. (2.14), we can use the latter to replace $M_1$ in computing $\{M_1, H\}$ which then is seen to vanish, thus the invariance of $M_1$ on this torus, because the other three integrals mutually Poisson commute. Similarly, $M_1$ Poisson commutes with $M_0$ and $M_{-1}$ within the torus (but not outside.) It can be checked that, in general, without the constraint of Eq. (2.14), $M_1$ is not invariant.

From the above explanation, we see that the "test" functional $\mathcal{F}$ used in Ref. [21] to establish the persistence criteria has some similarity with our torus-specific invariant but is obviously of different nature, for not used for defining the tori and for the requirement to Poisson commute with the other three functionals.

*2.3. Towards bridging integrability and nonintegrability*

Replacing $M_1$ in Eq. (2.14) by another $M_\tau$ with $\tau > 1$, not necessarily that defined by Eq. (2.13), we specify a different torus. The procedure can be continued. The tori may have intersection(s) on which the relevant $M_\tau$s mutually Poisson commute, as can be verified.

On the other hand, we can also replace $\lambda_{-1}M_{-1} + \lambda_0 M_0 + \lambda_1 M_1$ in Eq. (2.14) by $M = \sum_\tau \lambda_\tau M_\tau$, with $\tau$ not running over $-1$, 0 and 1, for different special solutions. $M_\tau$s, such as those defined by Eq. (2.13) for $\tau \geq 1$, however do not necessarily mutually Poisson commute with each other, neither with $H$. But, when we have $2K+1$ $M_\tau$s altogether, each Poisson commuting with $H$ (thus conserved) *on some specific torus*, we may term a kind of *pseudo-integrability* of the system: in the generalized notion of Vittot [37], the Hamiltonian is not *nonresonant* with such $M_\tau$s not mutually Poisson commuting on the corresponding torus.

The above discussion is quite general but does not appear to be connected with our focus on the longulent states of GrNLS closely enough. So, alternatively, we may define *pseudo-integrablity* more practically and more meaningfully here, with association to the longulent state and the corresponding whiskered torus, in the sense of specifying the latter with the "right" $M_\tau$s of a total number $\leq 2K+1$. Such a notion seems to be potential to lead to deeper mathematical developments, although no sufficient progresses have been made so far to offer a complete theory of longulence. We will come back to this after the analysis of the longulence developed from the multi-phase exact soluitons constructed with additional on-torus invariant, followed by Conjecture 1 with the associated remark on *a-posteriori* KAM theorem.

## 3. Longulence

As mentioned, Eq. (2.7) of the monochromatic wave or condensate solves also NLS, which means that the truncation is not relevant for the solution itself. However, with (modulational) instability, the final states should depend on the truncation threshold $K$. If the solution for the NLS is well-behaved (with no singular behaviors such as clapse or blow-up, say), then the GrNLS solution should converge, for $K \to \infty$, to that of NLS. Also, with an additional well-designed potential, as commonly in the so-called GP equation of BEC, convergence of GrNLS to NLS can also happen, which is the case in the work of Bland et al. [20] for the defocusing/repulsive case where the Galerkin approximation with the harmonic-oscillator eigenmodes rather than our Fourier modes was used: not surprisingly, they found dynamics close to full GP, which is similar to one of our cases below. In general, the focusing (Gr)NLS, like (Gr)BH and (Gr)CP in Ref. [16], do not have such convergence. Actually, as already indicated, our analyses and results apply also to the no-linear-dispersion regularization case, i.e., GrNLDNLS, thus nonconvergence of the corresponding Gr-system to the full-system in general (but not for the condensates). Working nevertheless with finite $K$, we will not consider the issue of convergence any more except for one apparent case.

*3.1. Numerical method: (pseudo-)spectral method aligning precisely with the Glerkin regularization*

The numerical analysis uses the standard (pseudo-)spectral method. To clarify its particular relation with our theoretical model in terms of Fourier Galerkin truncation, with still explain it briefly in the following, for the case of GrNLS.

Let the periodic lattice coordinate satisfy $x_j = x_{j+N}$, whence $\Psi(x_{j+N}) = \Psi(x_j) =: \Psi_j$ for $j = 0, 1, 2, ..., N-1$, defining a discrete torus $\mathbb{T}_N$. The theoretical foundation of the standard (pseudo-)spectral method and the lattice representation of the Gr-continuum lies in replacing $\hat{\Psi}_k$ defined by the *discrete Fourier transform* (DFT) for $|k| \leq M$ (with $N-1 = 2M$ here), $\hat{\tilde{\Psi}}_k := \sum_{x_j \in \mathbb{T}_N} \frac{\Psi_j}{N} e^{-\hat{i}kx_j} = \hat{\Psi}_k + \sum_{i \neq 0} \hat{\Psi}_{k+iN}$. The aliasing error, represented by the second term, can be mitigated using dealiasing techniques like zero-padding or, alternatively speaking, truncation at $K < N/4$ ("2/4-rule" for cubic nonlinearity, rather than the "2/3-rule" for quadratic nonlinearity). Unifying the dealiasing and the Galerkin truncation results in, correspondingly, $\hat{\hat{\psi}}_k = \hat{\psi}_k$ for $\psi = P_K \Psi$ in the GrNLS Eqs. (2.3, 2.4 and 2.5).

The 2/4-rule ensures sufficient sampling with $N$ sites for the Gr-continuum of $2K+1$ degrees of freedom, but in our numerical computations we typically have $N$ much larger than $4K$ to have smoother output of the results; for example, we have $N \geq 128$ for $K = 9$. The pseudo-spectral method marches in Fourier space, evaluating the nonlinear term in physical space via DFT of $P_K 2|\psi_j|^2\psi$. So, the computation method aligns precisely with the Gr-systems, with only errors from the computer roundoff and time discretization (the standard fourth-order Runge-Kutta scheme).

Other appropriate computational methods in principle can also reproduce the main features of our numerical results, with less accuracy and requiring more carefullness though.

### *3.2. GrNLS longulence*

#### *3.2.1. GrNLS longulence from the travelling-wave initial data*

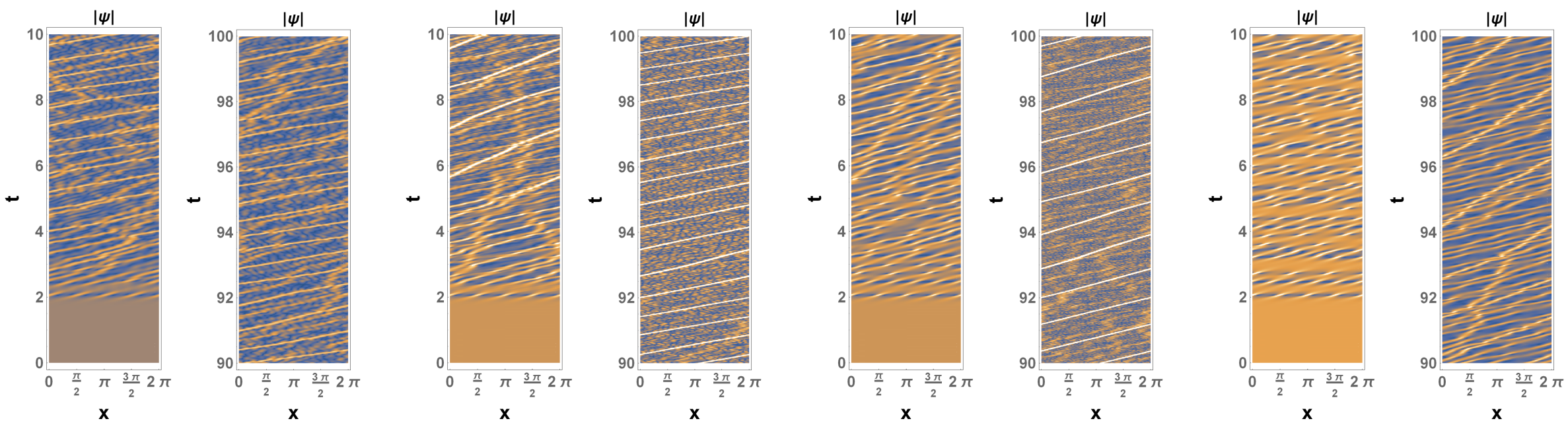

(a) $K = 9$. (b) $K = 15$. (c) $K = 24$. (d) $K = 33$.

Figure 1: The carpets/contours of the GrNLS fields starting from Eq. (2.7) at $t = 0$, for a travelling wave with $\alpha = 1$, $S = 3$: lighter colors indicate larger values in all figures, coded per panel (always for all figures).

We start the numerical analysis of the development of GrNLS longulence using the travelling-wave initial data: Fig. 1 shows the evolution into a stable (statistically, with respect to the small scale weaker disordered component) pseudo-periodic state (after around $t = 2$) with longons, which is the case even for $K = 1$ (not shown). Different $S$ and $K$ lead to quantitatively different patterns in the sub-figures, but with the same qualitative scenario characterized by solitonic longons admist disordered components, appearing to converge to NLS dynamics. The results indicate smaller speeds of the solitonic longons for larger $K$s.

#### *3.2.2. GrNLS longulence from the two-phase-solution initial data*

We then turn to the development of longulence from the two-phase-solution initial data, as given by Fig. 2, including the real part of $\psi$ for visualization of the character of the early exact solution. The scenario is similar, with no essential differences between periodic and quasi-periodic cases, satisfying the Diophantine condition or not (thus other parameterizations with, say, $\lambda_0 = 13 + \sqrt{2}$ and $\lambda = 2/3$ are not shown).

Removing $S^2$ under $\sqrt{\ }$ in Eq. (2.11), we have the corresponding GrNLDNLS solution at $t = 0$ of which is used as the initial data for the development of the longulence shown in Fig. 3: the contours and profiles show that, although the "solitization" is not strong enough to have as marked longons with clear periodicity as in other cases, non-thermalization is obvious for no homogenization of the pulses into "noise" (with the peaks and dips of $|\psi|$ respectively developed "further" rather than "closer").

#### *3.2.3. GrNLS longulence from the three-phase-solution initial data*

Finally, we examine the case with the initial data corresponding to the three-phase solutions specified by an additional on-torus invariant. With slightly more careful observation, we still see similar solitonic longons in the well-developed stage in the right panel of Fig. 4; the latter appears to indicate that the well-developed longulence is closer to the exact solution, compared to Figs. 1 and 2, as may be observed from the patterns.

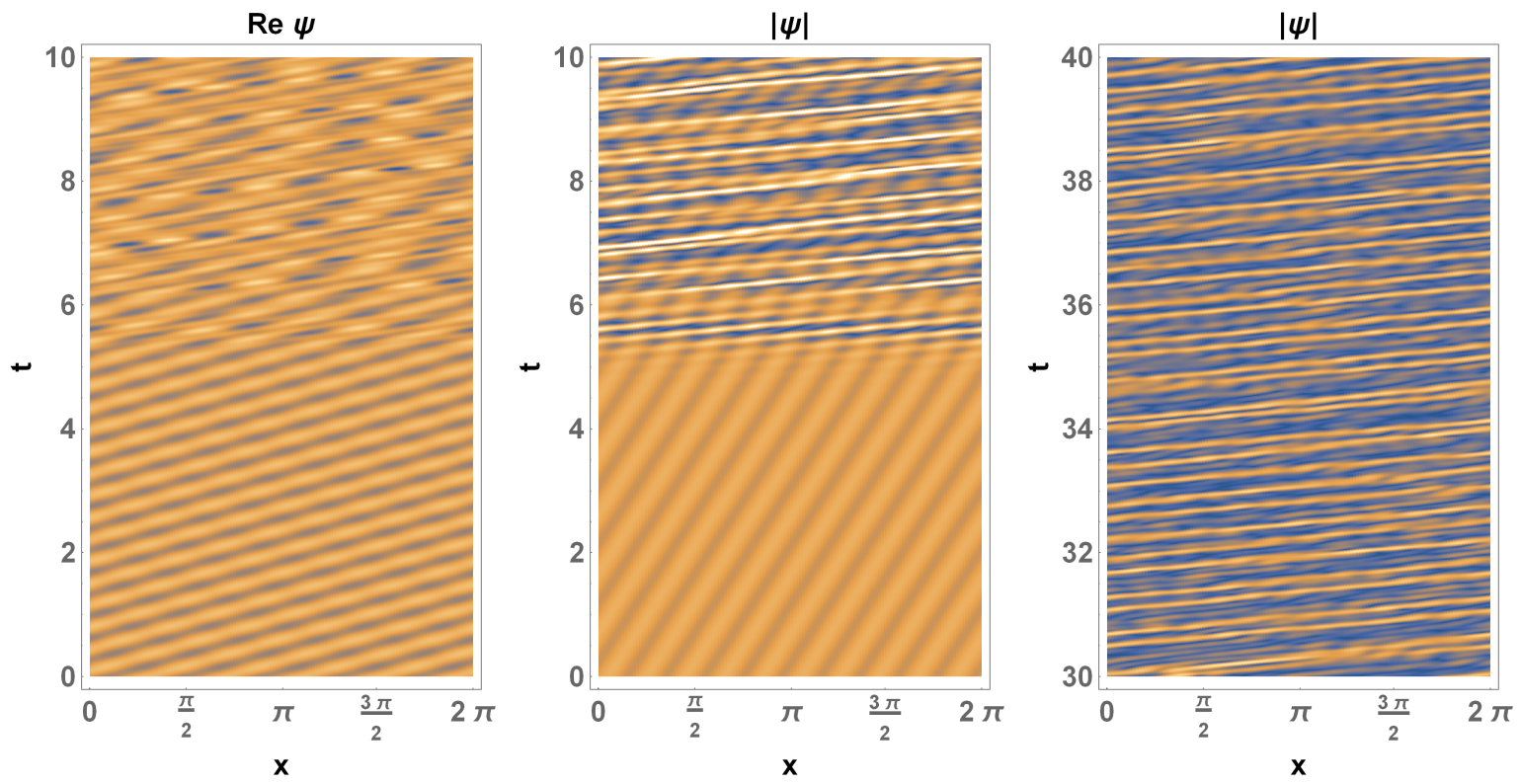

Figure 2: The carpets/contours of the GrNLS fields starting from the two-frequency quasi-periodic solution data given by Eq. (2.11) at $t = 0$, with $S = 5$, $K = 9$, $\lambda_0 = 11.3 + \pi$ and $\lambda_{-1} = \frac{\sqrt{6}}{3.7}$.

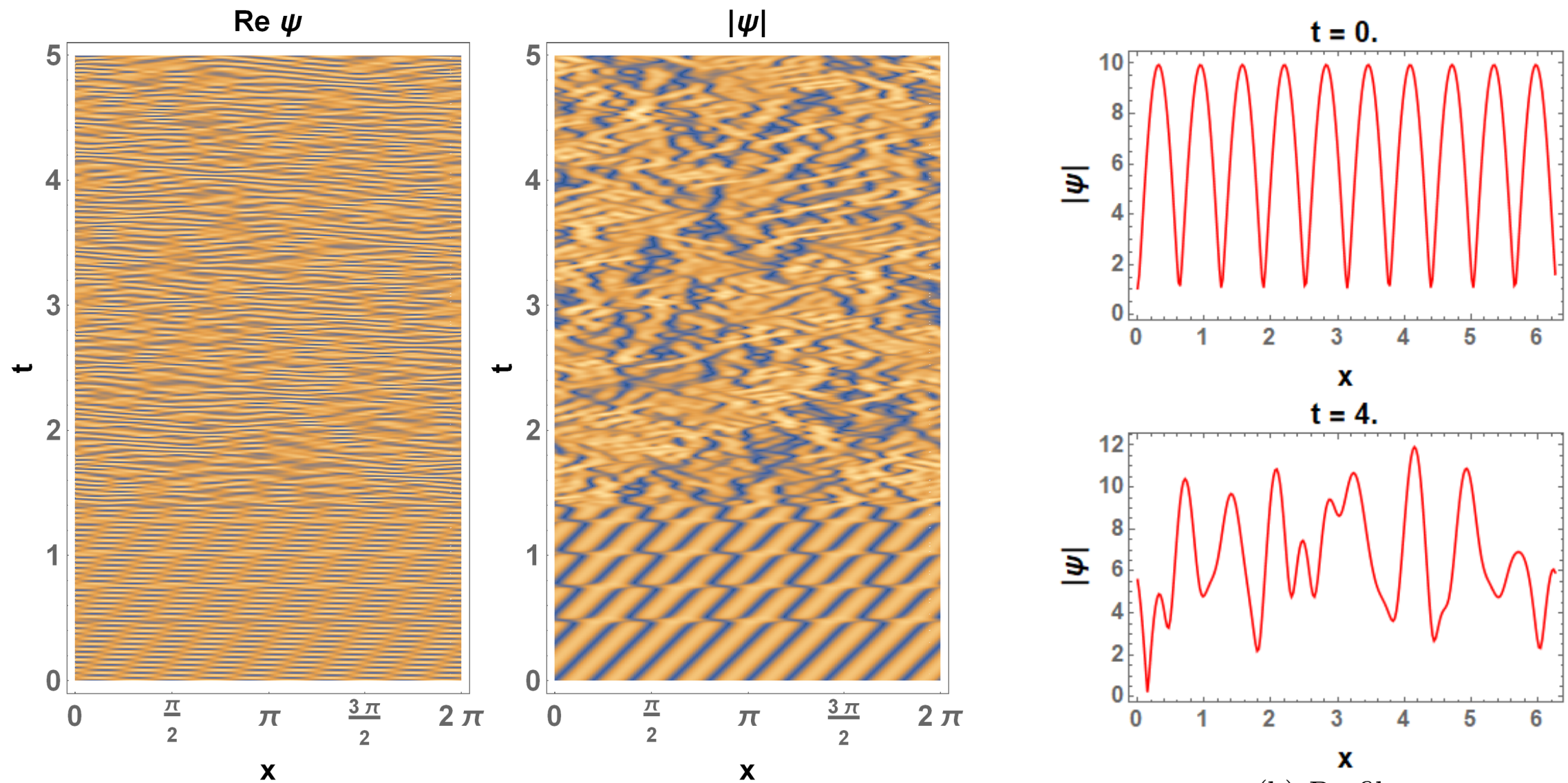

(a) Contours. (b) Profiles.

Figure 3: GrNLDNLS two-frequency case, corresponding to that of Fig. 2, with $S = 5$, $K = 9$, $\lambda_0 = 10\sqrt{2}/1.414$ and $\lambda_{-1} = -10\pi$.

Removing $S^2$ under $\sqrt{\ }$ in Eq. (2.15), we have the corresponding GrNLDNLS solution at $t = 0$ of which is used as the initial data for the development of the longulence, with features similar to the case given by Fig. 3, thus not shown.

#### *3.2.4. Pseudo-integrability conjecture*

All the numerical results indicate a kind of universal attractor characterized by solitonic longons among less-ordered components, which should be underlined by some high-dimensional whiskered tori, but some details and points should be emphasized and further clarified, thus the following extended remarks.

Even the quasi-periodic solutions with rugged invariants are not sufficient to have enough stability, resulting in far different longulent states, which may be a further indication of the relevance of additional on-torus invariants. So, as tried in Fig. 4, it is possible to similarly construct invariant tori with much closer longulence developed. In all our numerical tests of multi-frequency tori, except for minor "improvements" concerning stability, rational independence or the Diophantine condition of the frequencies does not appear to have essential effects on the developments of longulence (thus other similar results corresponding to Fig. 4 not shown). The indication may be that the (universal) longulent state is the only stable attractor or that the initial datum prepared as such is close enough, but not precisely the right corresponding whiskered torus.

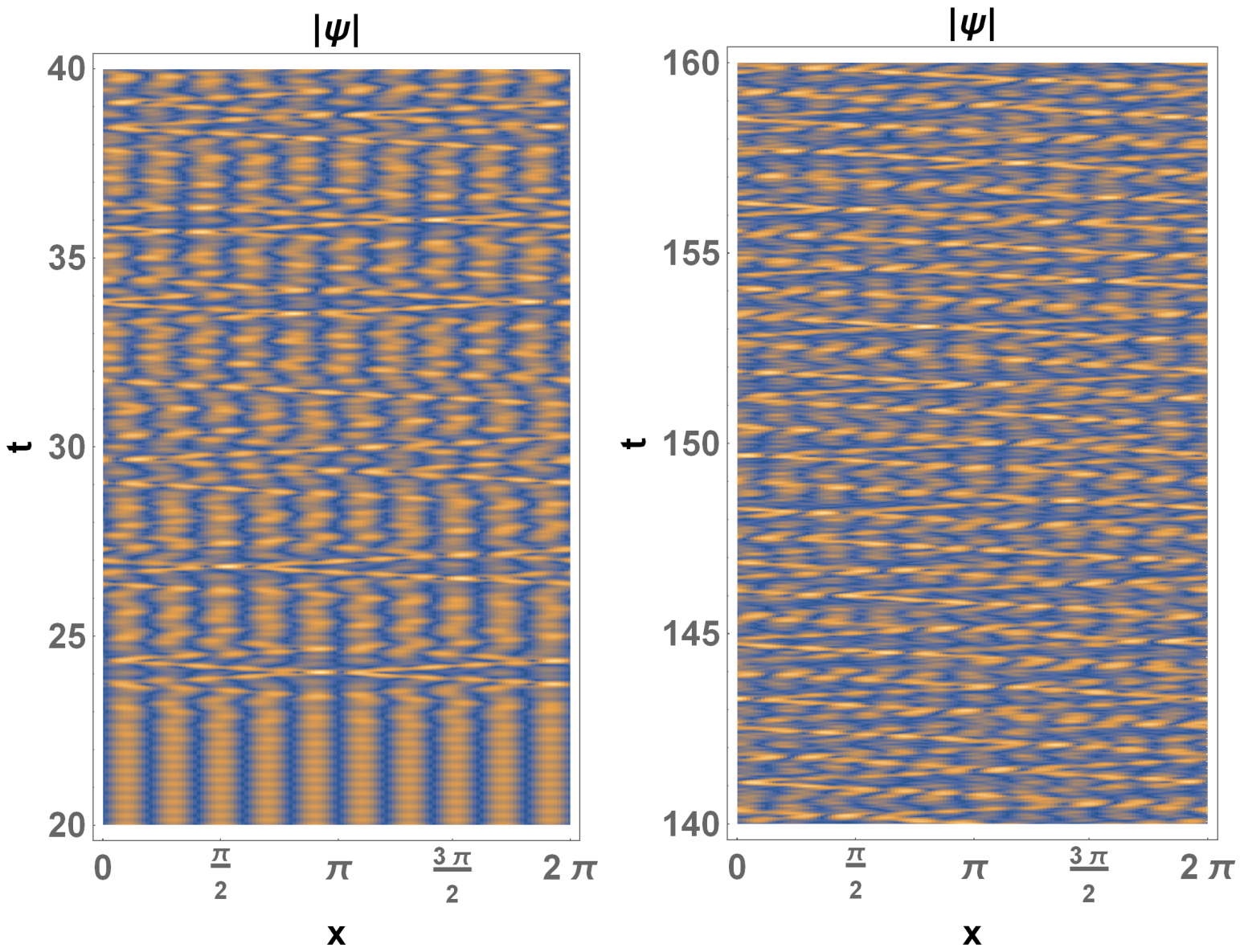

Figure 4: The carpets/contours of the GrNLS fields starting from the three-frequency-solution initial data given by Eq. (2.15) at $t = 0$, with $\theta_0 = \pi/2$, $S = 5$, $K = 9$, $\lambda_{-1} = \lambda_1 = -1$ and $\lambda_0 = -2.1\lambda_{-1}S$.

For given parameters such as $S$ and $K$, there is of course the problem of what the better choice of the on-torus invariants is to have the (unstable) solution closer to final longulent state. Systematic improvements might be made by the choice of more appropriate torus-specific invariants, but we so far do not really have a good theory, for lack of a mathematical structure, say, as that for the NLS [1]. We have the following

**Conjecture 1.** *There are "right" torus-specific invariants supporting "pseudo-integrability", in the sense of specifying precisely a whiskered torus responsible for the longulent state.*

Together with the cases in Ref. [16], it also appears possible to have a unified, rather than case-by-case, proof of the corresponding whiskered torus with the *a-posteriori* KAM scheme, and the exact solutions (as constructed here) from which the longulent state develops might be used as the starting approximation.

### *3.3. No GrCGL nontrivial longulence?*

A natural further question is the (non)persistence of the (solitonic) GrNLS longons to GrCGL perturbation, which however appears even more difficult than those associated to the NLS [38, 21]. A relatively simpler way to have the clear persistence result is to replace $\partial_{xx}\Psi$ by some appropriate (pseudo-)differential operator on $\Psi$, resulting in some kind of hyper-dispersion and hyper-dissipation with a sharp transition across $K$ for the dissipation and dispersion functions, and to add an independent driving such as $\epsilon\psi$ show-cased for GrBH in Ref. [16]. But, here, rather than getting around the difficulties, we tentatively analyze some numerical experiments performed in the standard (Gr)CGL framework, aming to offer clues and for motivation.

For the CGL equation,

$$\hat{i}\partial_t\Phi + C\partial_{xx}\Phi + 2G|\Phi|^2\Phi = 0 \tag{3.1}$$

where $C = 1 + \hat{i}\eta$ and $G = 1 + \hat{i}\epsilon$, with possibly an addtional term to be picked up later. Like NLS, the addition of the corresponding truncation force $g$ turns CGL into GrCGL for $\phi$.

The GrCGL dynamics occupying a single wavenumber $k = S$ reads

$$\hat{i}\dot{\hat{\phi}}_S = CS^2\hat{\phi}_S + 2G|\hat{\phi}_S|^2\hat{\phi}_S, \tag{3.2}$$

and the solution similar to Eq. (2.8) for GrNLS is

$$\phi = \alpha S e^{\hat{i}S[x-(C-2\alpha^2 G)St]}. \tag{3.3}$$

When $Im(C) = 2\alpha^2 Im(G)$, or $\eta/\epsilon = 2\alpha^2$, we have the travelling-wave solution with wave speed $c = [Re(C) - 2\alpha^2 Re(G)]S$ or $c = (1 - 2\alpha^2)S$ as for the GrNLS. Again, this solution also solves the untruncated CGL. No nontrivial (statistically) stable GrCGL longulent states have been found in our numerical experiments starting from this solution, with everything else the same as in the GrNLS case in Fig. 1.

We have so far neither been able to generate GrCGL longulent states from the (quasi-)periodic (2.11) and (2.15).

As mentioned, it is possible that with more appropriate GrNLS torus-specific invariants, we could have the final longulence very close to the initial one; then, it could be that persistence to the GrCGL perturbation takes place under suitable conditions. The difficulties associated to issues such as the nontrivial pseudo-differential operator of the truncation and the numerical subtleties for extremely small $\eta$ and $\epsilon$ however have hindered good progress.

From weakly-nonlinear dynamics point of view, it appears important to have the linear damping or forcing be appropriately balanced. With constant $r$ for such an additonal perturbation term $\hat{i}r\Phi$, thus balance formally taking place only at a single scale (wavelength), it is possible but nontrivial for the existence of (quasi-)periodic solutions, since multi-scales are excited by nonlinearity. The necessity of the linear term is the usual case in physics [22], and, in mathematical treatments only special choice of the parameters have been found possible to have the CGL quasi-periodic solutions [21, 39, 23, 24].

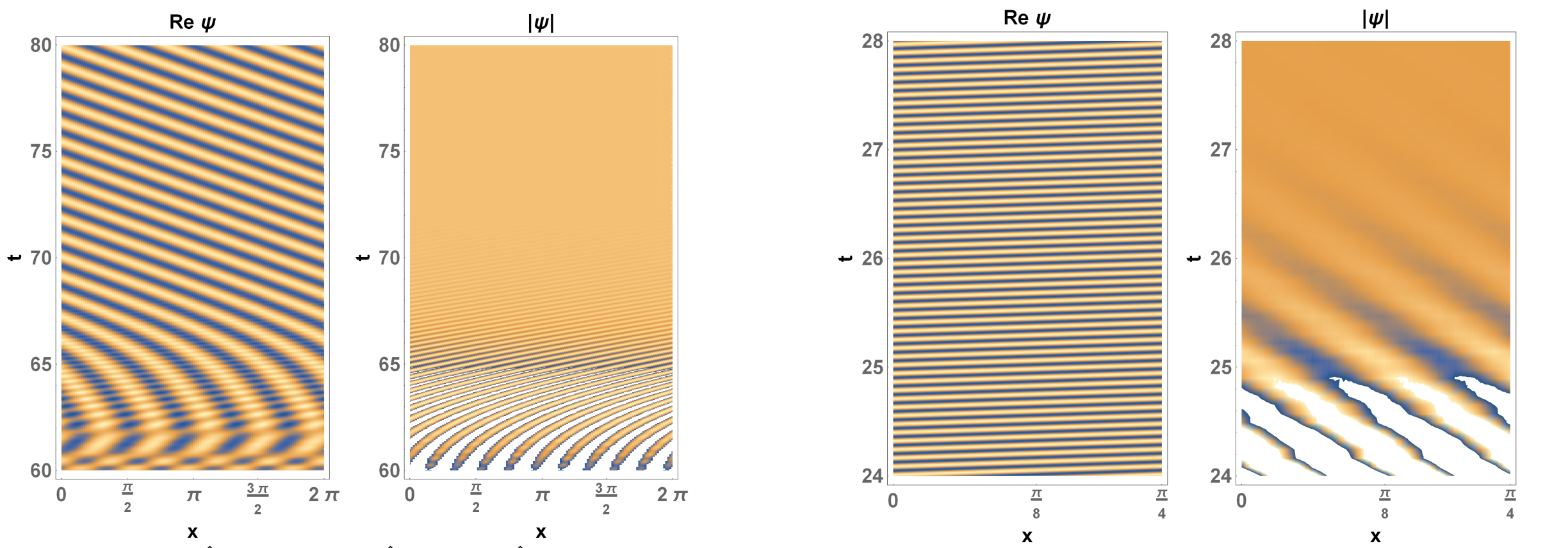

(a) $C = 1+0.04\hat{i}$, $G = 1+0.02\hat{i}$, $r = 0.04\hat{i}$, $S = 3$ and $K = 5$.

(b) $C = 1 + 0.02\hat{i}$, $G = 1 + 0.01\hat{i}$, $r = 0$, $S = 5$ and $K = 9$.

Figure 5: GrCGL patterns starting from Eq. (2.15) for exact three-phase NLS solutions, with $\theta_0 = \pi/2$, $\lambda_{-1} = \lambda_1 = -1$, $\lambda_0 = -2.1\lambda_{-1}S$.

For GrCGL, the corresponding $\hat{i}r\phi$ then may not be necessary, since the Galerkin regularization term $g$ can make the nonlinearity strong, thus even more nontrivial balance could happen. When the inertial manifold property is in control and the truncation wavenumber is large, things become kind of trivial because of the convergence to the full CGL, in which case the truncation effect is a small perturbation.

We found in various numerical tests the final "clean" periodic solutions, in abscence of any signature of disorder, with and without the $\hat{i}r\phi$ term from the corresponding cases with everything else the same as used for GrNLS. For example, Fig. 5, for cases corresponding to Fig. 4, shows the evolution to constant-amplitude travelling-wave solutions. The travelling waves are of shortest respective wavelengths, i.e., $|k| = K$ and of the form (3.3) found earlier (verified by checking the relations between the respective wave amplitudes and speeds) but not of the linearized dynamics. [Plotting only for the smaller region in the case of $S = 5$ and $K = 9$ is to avoid the artificial Moré patterns.] Note that, unlike the KAM results in Refs. [23, 24], we did not start from the solutions of the linearized system, thus not of that perturbative nature.

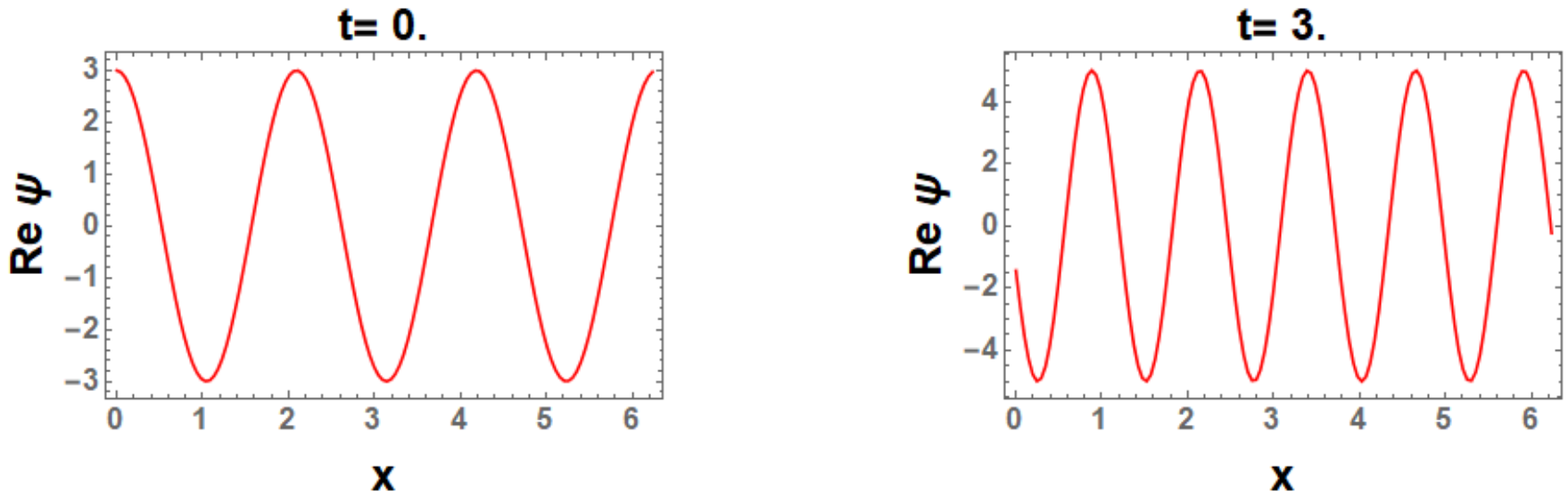

Figure 6: The GrCGL transition of condensate from $k = S$, for initial data (left) prepared with $\alpha = 1$, $C = 1 + 2\hat{\imath}$, $G = 1 + \hat{\imath}$, $r = 0$, $S = 3$ in Eq. (3.3), to a final condensate wave profile of $k = K = 5$ (right).

Results similar to Fig. 5 for GrCGL with (much) larger $\eta$, $\epsilon$, with or without (large) $r$, are also found, with faster convergence. For example, Fig. 6 presents the transition of condensate from $k = S$ to $k = K$ purely by instability, subjecting only to roundoff error perturbation as in all previous cases, in the pseudo-spectral computation, with the final state return to the monochromatic wave (3.3) of the same $\alpha = 1$ but of different wave number. [Periodic solutions of such large $K$s were never reported, to the best of our knowledge (for instance, Ref. [30] studied bifurcation and chaos of low-dimensional dynamics).]

More results, such as the convergence to the condensate at $K$ with $\alpha \neq 1$, with other initial data, have also been collected. No satisfying theory for these observations are available so far (see below). Having not found quasi-periodic or longulent GrCGL states though, we have no reason to exclude the existence. Combing with the chaotic aspect ([30] and references therein), we still tend to believe that nontrivial GrCGL longulent states are still possible, deserving further remarks.

## 4. Expectation

Good understanding of relevant GrCGL dynamics beyond the small perturbation to GrNLS can be beneficial for the latter, by learning from the differences, say, as is the purpose here. So, we proceed by noting that the CGL damping and forcing may balance on particular orbits/tori in such a way that some on-torus invariants present and longulence emerges.

Although explicit multi-frequency CGL solutions have not been found analytically, there are suggestions of the existence of quasi-periodic (whiskered) tori [21, 39, 23, 24]. Note that the formulas from the KAM method [Eqs. (6) and (3.22) of, respectively, Refs. [23, 24]] are with whiskered components, qualitative or asymptotic, and, are only for particular choices of parameters. What's more, the term proportional to $\Phi$ (also physically important [22]) is crucial in their KAM method, although it balances only one scale. In the GrCGL case with truncation $g$, this term however appears not as needed. Nevertheless, we have not yet been able to construct GrCGL nontrivial longulent states from the corresponding GrNLS data.

New techniques are needed. The Lyapunov-function approach seems promising, with however caveats: for instance, obviously, $\int u^{2n} dx$ for any integer $n > 0$ is a Lyapunov function for $u_t - u^2 u_x = -u$ controlling real $u$, and the pattern selection was not found to minimize the one written down in Ref. [40] for the slightly more complex dynamics.

As already mentioned, the on-torus-invariant effort shares some similarity with part of the Melnikov method used in Ref. [21], and we may hopefully expect further combination for more powerful techniques. For general (Gr)CGL with coefficients neither appropriate for a perturbative treatment [21] nor so special to have the nearly (quasi-)periodic solution for the corresponding linearized system [23], it is not impossible still to simultaneously set up the right multiple on-torus invariants, probably correspondingly an appropriate Lyapunov function, to construct specific invariant (whiskered) tori for nontrivial longulence.

If the GrCGL quasi-periodic or longulent states can be found as for GrNLS and other hydrodynamic-type Gr-systems [16], then we are closer to the *a-posteriori* KAM scheme that would assure the existence of (whiskered) tori close by (e.g., de la Llave and collaborators' recent works, including Ref. [41] on partial differential equation and Ref. [42] on maps, and references therein). The hydrodynamic-type Gr-systems studied in Ref. [16] present persistence of those longulent states against appropriate (small) forcing and damping, which of course carries over to Gr(NLD)NLS, *mutatis mutandis*, may be regarded as the support of the existence of nontrivial GrCGL longulence, more subtle though.

## Acknowledgement

The author is indebted to D. Levermore for discussing Ref. [21].